\documentstyle[preprint,tighten,aps,prd]{revtex}
\begin{document}
\draft
\preprint{}
\newcommand{\beq}{\begin{equation}}
\newcommand{\eeq}{\end{equation}}
\newcommand{\beqa}{\begin{eqnarray}}
\newcommand{\eeqa}{\end{eqnarray}}
\newcommand{\bdm}{\begin{displaymath}}
\newcommand{\edm}{\end{displaymath}}

\title{Dirac versus reduced quantization and operator ordering}
\author{K Shimizu\thanks{e-mail address: shimizu@u-aizu.ac.jp}}
\address{Center for mathematical sciences,
 The University of Aizu,
 Aizu-Wakamatsu City,
 Fukushima, 965 Japan}
\maketitle 

\begin{abstract}
We show an equivalence between Dirac quantization and the reduced phase space
quantization. The equivalence of the both quantization methods determines the 
operator ordering of the Hamiltonian. Some examples of the operator ordering
 are shown in simple models.\\
\end{abstract}

\section{Introduction}
  In recent years some authors\cite{kun} are discussing on Dirac 
quantization and the reduced phase space quantization. Their arguments are
 that the reduced phase space quantization and Dirac quantization may be
 different in the constraint system with a non-trivial metric.  In order
 to clarify the problem, let us consider the simplest model, as an
 example.\par
  Lagrangian is given by
\beq
 L=\frac{1}{2}\dot{x}^2+\frac{f(x)}{2}(\dot{y}-\lambda)^2 \label{eqn:l1}
\eeq
where $\lambda$ is a Lagrange multiplier. There is a non-trivial metric
 $f(x)$.  This is not a field theory but a
 quantum mechanics. The Hamiltonian of this system is
\beq 
   H=\frac{1}{2}p_{x}^{2}+\frac{1}{2f(x)}p_{y}^{2}+\lambda p_y
   \label{eqn:a1}   
\eeq
and there are two constraints
\beq
                 p_\lambda\equiv\pi\approx 0,
\eeq
\beq
                 p_y\approx 0.
\eeq
These are first-class constrains. We set $p_y=0$ in the Hamiltonian before
 the quantization. Then the Hamiltonian reduces to
\beq 
                 H=\frac{1}{2}p_{x}^{2}                          
\eeq
and the Hamiltonian operator is
\beq
              \hat{H}=\frac{1}{2}\partial_{x}^{2}. \label{eqn:a2}
\eeq
This is the reduced phase space quantization. The procedure of the reduced
phase space quantization is to reduce first and then quantize.\par
 In the case of Dirac quantization, its procedure is to quantize first and
then reduce. The Hamiltonian in this model is defined on the
 two-dimensional space of $x$ and $y$ without a constraint term. To
 ensure the invariance under the coordinate transformation, the
 Hamiltonian operator is written by
\beq
       \hat{H}=\frac{1}{2\sqrt{f}}\partial_x \sqrt{f}\partial_x+
             \frac{1}{2\sqrt{f}}\partial_y\sqrt{f}\frac{1}{f}\partial_y,
\eeq
where $\sqrt{f}$ is $\sqrt{det g_{\mu\nu}}$. The metric $g_{\mu\nu}$ is the
two-dimensional metric of $x$-$y$ space. Since $\hat{p}_y=\partial_y\approx 0,
 y$ derivatives in the Hamiltonian operator 
are eliminated. Then the Hamiltonian operator in Dirac quantization is
\beq
        \hat{H}=\frac{1}{2\sqrt{f}}\partial_x\sqrt{f}\partial_x.
\eeq
This is not same with the result of the reduced phase space quantization.
 This is the problem of an inconsistency of the reduced phase space
 quantization and Dirac quantization.\par
 In section 2 we show the equivalence of the both quantization methods. 
It is shown that the Hamiltonian operator of Dirac quantization should
include the constraint term and be invariant under the three-dimensional
coordinate transformation of $x,y$, and a configuration variable conjugate
to the Lagrange multiplier.\par
 In section 3 we discuss a problem of the operator ordering. If the
 Hamiltonian has a non-trivial metric;
\beq
          H=\frac{1}{2}g^{\mu\nu}p_\mu p_\nu,
\eeq
the Hamiltonian operator may have a function of scalars like$ R,
 R^{\mu\nu}R_{\mu\nu}, R^{\mu\nu\lambda\sigma}R_{\mu\nu\lambda\sigma},
\cdots$ from the invariance of the coordinate transformation, in addition
 to the Laplacian
\beq
       \hat{H}=\frac{1}{2}\bigtriangleup + F( R, R^{\mu\nu}R_{\mu\nu},
 R^{\mu\nu\lambda\sigma}R_{\mu\nu\lambda\sigma},\cdots ).
\eeq
The Laplacian is indispensable from the invarianse and derived from the
 descretized path integral\cite{shi}. The additional function is called
the quantum mechanical potential. The problem of operator ordering is
 to determine the quantum mechanical potential. Using the equivalence
 between the reduced phase quantization and Dirac quantization, we
 determine this potential in simple models.\par
 Section 4 is devoted to the conclusions and discussions.

\section{The reduced phase space quantization and Dirac quantization}
Let us reconsider the Hamiltonian (\ref{eqn:a1}). We take a gauge
 condition $\dot{\lambda}$=0 to quantize this system in the path integral
formalism.  The Hamiltonian form path integral\cite{fra} is given by
\beqa
   Z&=&\int d\mu exp[iS],\nonumber\\ 
   d\mu&=&[dxdp_xdydp_yd\pi d\lambda],\nonumber\\
   S&=&\int dt p_x\dot{x}+p_y\dot{y}-\pi\dot{\lambda} 
    -\frac{1}{2}p_{x}^{2}-\frac{1}{2f(x)}p_{y}^{2}-\lambda p_y.
    \label{eqn:p1}
\eeqa
$\lambda$ is a momentum variable so that the sign of gauge fixing term is
negative. After the partial integration, it becomes usual one.  Since this
 gauge is an Abelian, we need not to introduce any ghost. After
the integration of $\pi, \lambda,p_y$, and $y$, the partition function 
becomes
\beqa
    Z&=&\int d\mu' exp[iS'],\nonumber\\ 
    d\mu'&=&[dxdp_x],\nonumber\\ 
   S'&=& \int dt p_x\dot{x}-\frac{1}{2}p_x^2. 
     \label{eqn:p2}
\eeqa
This is nothing but the partition function of a free particle. Then the
 Hamiltonian operator is equation (\ref{eqn:a2}). This means the operator
 formalism corresponding to the stage of the path integral (\ref{eqn:p2})
is the reduced phase quantization. On the other hand, Dirac quantization
is the operator formalism corresponding to the stage of the path
 integral (\ref{eqn:p1}). In equation (\ref{eqn:p1}) any variable is not
 integrated and constraint variables are still alive.
The symmetry of this path integral is the coordinate transformation of
the whole configuration space including $\pi$ which is a configuration 
variable conjugate to $\lambda$. Therefore, the Hamiltonian operator should
be made invariant under the three-dimensional coordinate transformation,
not the two-dimensional one. Then the Hamiltonian operator is
\beqa
   \hat{H}&=&\frac{1}{2\sqrt{g}}\partial_\mu \sqrt{g}g^{\mu\nu}
                                 \partial_\nu\nonumber\\
          &=&\frac{1}{2}\partial_x^2+\frac{1}{2}\partial_y f\partial_y
          +\frac{1}{2}\partial_y\partial_\pi +\frac{1}{2}\partial_\pi\partial_y
\eeqa
where $g_{\mu\nu}$ is a inverse of $g^{\mu\nu}$of the Hamiltonian. The
original Lagrangian (\ref{eqn:l1}) has a singular metric. However,
the gauge fixed Lagrangian which is made by the integration of momentum
variables in equation (\ref{eqn:p1}) has a regular metric and it 
coincides with the inverse of $g^{\mu\nu}$.\par 
     Using the constraint $\hat{p_y}=\partial_y\approx 0$, we get the
 Hamiltonian
\beq
    \hat{H}=\frac{1}{2}\partial_x^2.
\eeq
This is Dirac quantization and we obtain the same Hamiltonian operator
 with the reduced phase space quantization. This is natural because we 
start from the same path integral (\ref{eqn:p1}). This simplest example 
indicates that Dirac quantization and the reduced phase space quantization
should be coincide.\par
A naive Dirac quantization showed in the introduction is made by the
requirement that the Hamiltonian operator should be invariant under the
coordinate transformation of $x$ and $y$. In that case the constraint
term is treated separately. However, under some coordinate transformation,
the net Hamiltonian and the constraint term are mixed. The naive Dirac
quantization does not represent the symmetry correctly. This is the reason
why the naive Dirac quantization is different from the reduced phase space
quantization.\par
  In general case with many variables, we can propose Dirac quantization 
and the reduced phase space quantization are equivalent because both 
quantizations are the operator versions of the different forms of the same
 path integral as before. We can determine the quantum potentials with
 this property.

\section{The operator ordering}
Let us now consider the Lagrangian
\beq
 L=\frac{1}{2}h(x)\dot{x}^2+\frac{g(x)^2}{2f(x)}(\dot{y}
-\frac{\lambda}{g(x)})^2. \label{eqn:l2}
\eeq
This Lagrangian leads the Hamiltonian
\beq
   H=\frac{1}{2h(x)}p_{x}^{2}+\frac{f(x)}{2g(x)^2}p_{y}^{2}+\frac{1}{g(x)}
\lambda p_y,\label{eqn:h1}
\eeq
and constraints
\beqa
 p_y\approx 0,\nonumber\\
 p_\lambda\equiv\pi\approx 0 ,
\eeqa
 as before. The reduced phase space quantization makes the Hamiltonian
\beq
        H=\frac{1}{2h(x)}p_x^2
\eeq
by the constraints. Then the Hamiltonian operator is
\beq
       \hat{H}=\frac{1}{2\sqrt{h}}\partial_x\frac{1}{\sqrt{h}}\partial_x.
\eeq
While in Dirac quantization, we consider the Hamiltonian in three-dimension
at first. The invariance of the three-dimensional coordinate transformation
allows the Hamiltonian operator of the form
\beqa
       \hat{H}&=&\frac{1}{2}\bigtriangleup + F( R, R^{\mu\nu}R_{\mu\nu},
 R^{\mu\nu\lambda\sigma}R_{\mu\nu\lambda\sigma},\cdots )\nonumber\\
&=&\frac{1}{2g\sqrt{h}}\partial_x\frac{g}{\sqrt{h}}\partial_x
  +\frac{1}{2g\sqrt{h}}\partial_y\frac{f\sqrt{h}}{g}\partial_y
  +\frac{1}{2g\sqrt{h}}\partial_y\sqrt{h}\partial_\pi \nonumber\\
 & +&\frac{1}{2g\sqrt{h}}\partial_\pi\sqrt{h}\partial_y
  +F
\eeqa
because in this model $R,\cdots$ are not zeros. $g_{\mu\nu}$ is an inverse
matrix of $g^{\mu\nu}$ of the Hamiltonian and is same with that of the
gauge fixed Lagrangian as before. The constraint $\hat{p_y}=\partial_y
\approx 0$ makes the Hamiltonian simple form; 
\beq
    \hat{H}=\frac{1}{2g\sqrt{h}}\partial_x\frac{g}{\sqrt{h}}\partial_x +F.
\eeq
     \par
    The inner product for the reduced phase quantization is defined as
\beq
             \int \sqrt{h}dx\Psi_r^* \Psi_r
\eeq
where $\Psi_r$ is a wave function of the reduced phase quantization. On 
the other hand, for Dirac quantization, it is written by
\bdm
            \int \sqrt{h}g dxdyd\pi \Psi_D^*\Psi_D
\edm
where $\Psi_D$ is a Dirac quantized wave function. Since the constraint
$\pi\approx 0$ means $\int d\pi \Psi_D^*\pi\Psi_D=0$, $\Psi_D^*\Psi_D$ is
proportional to $\delta (\pi)$ and could be written as $\Psi_D'^*\Psi_D'
\delta(\pi)$. We rewrite $\Psi_D'$ as $ \Psi_D$ again
 and the inner product reads
\beq
           \int dy\int \sqrt{h}g dx \Psi_D^*\Psi_D.
\eeq
 $\int dy$ is a gauge volume and it should be ignored.  For the both inner
 products to agree with each other,
\beq
                    \Psi_D=\frac{1}{\sqrt{g}}\Psi_r
\eeq
must be satisfied. \par
 An expectation value of the energy for the reduced phase space quantization
is
\beqa
 <E>_r&=&\int \sqrt{h} dx \Psi_r^* \hat{H}\Psi_r\nonumber\\
    &=&\int \sqrt{h} dx \Psi_r^*\frac{1}{2\sqrt{h}}\partial_x
      \frac{1}{\sqrt{h}}\partial_x\Psi_r.
\eeqa
While in Dirac quantization it is given by
\beqa
 <E>_D&=&\int \sqrt{h}gdx \Psi_D^* \hat{H}\Psi_D\nonumber\\
    &=&\int \sqrt{h}gdx\frac{1}{\sqrt{g}}\Psi_r^*(
\frac{1}{2g\sqrt{h}}\partial_x\frac{g}{\sqrt{h}}\partial_x+F)
\frac{1}{\sqrt{g}}\Psi_r\nonumber\\
&=& <E>_r\nonumber\\
& +& \int\sqrt{h}dx\Psi_r^*(-\frac{g''}{4hg}
+\frac{g'^2}{8hg^2}+\frac{g'h'}{8gh^2}+F)\Psi_r,
\eeqa
where $'$ is a $x$ derivative. To be consistent with each other, the second
 term should be zero in the last equation. In other words, the function F is
 determined so that the both quantization methods coincide.\par
     In this space, $R$ and $R^{\mu\nu}R_{\mu\nu}$ are written as
\beq
        R=\frac{1}{h}(-\frac{g''}{g}+\frac{g'^2}{g^2}+\frac{g'h'}{2gh})
          -\frac{g''}{gh},
\eeq
\beq
       R^{\mu\nu}R_{\mu\nu}=(-\frac{g''}{g}+\frac{g'^2}{g^2}
     +\frac{g'h'}{2gh})^2\frac{1}{h^2}+\frac{1}{2}(\frac{g''}{gh})^2.
\eeq
If we define
\beq
     A\equiv\frac{1}{h}(-\frac{g''}{g}+\frac{g'^2}{g^2}+\frac{g'h'}{2gh}),
\eeq
\beq
          B\equiv\frac{g''}{gh},
\eeq 
$R$ and $R^{\mu\nu}R_{\mu\nu}$are rewritten as
\beq
           R=A-B,
\eeq
\beq
         R^{\mu\nu}R_{\mu\nu}=A^2+\frac{1}{2}B^2.
\eeq
From these equations, we get
\beq
            A=\frac{R\pm \sqrt{6R^{\mu\nu}R_{\mu\nu}-2R^2}}{3}.
\eeq
Then if we take
\beq
            F=\frac{-R+\sqrt{6R^{\mu\nu}R_{\mu\nu}-2R^2}}{12},
\eeq
$<E>_r$ coincides with $<E>_D.$ Here we take a positive sign of root. We 
discuss the reason later. The operator ordering for the 
Hamiltonian (\ref{eqn:h1}) is, then,
\beq
      \hat{H}=\frac{1}{2}\bigtriangleup+
            \frac{-R+\sqrt{6R^{\mu\nu}R_{\mu\nu}-2R^2}}{12}.
       \label{eqn:o1}
\eeq

  Let us consider the next example. The Hamiltonian is
\beqa
    H&=&\frac{1}{2}g^{\mu\nu}p_{\mu} p_{\nu}\nonumber\\
 &=&\frac{1}{2}\gamma^{ij}p_i p_j-\frac{N^i}{M}p_i p_y+\frac{N^2}{2M^2}
   p_y^2+\frac{1}{M}p_y\lambda,
     \label{eqn:h2}
\eeqa
where $p_i$ means $p_{x^i}$ and $i$ runs from 1 to n. $\lambda$ is a
 Lagrange multiplier. The metric $g^{\mu\nu}$ of the Hamiltonian and its
inverse which accords with the metric $g_{\mu\nu}$ of the gauge fixed 
Lagrangian are
\beq
 g^{\mu\nu}=\left(\begin{array}{ccc}
             \gamma^{ij} & -\frac{N^j}{M} & 0\\
            -\frac{N^i}{M} & \frac{N^2}{M^2} & \frac{1}{M}\\
                 0        & \frac{1}{M} & 0
           \end{array}
    \right),                  \label{eqn:g1}
\eeq
\beq
 g_{\mu\nu}=\left(\begin{array}{ccc}
             \gamma_{ij} & 0 & N_j\\
                 0        &  0 & M\\
            N_i & M & -N^2+N_i N_j
           \end{array}
   \right).                  \label{eqn:g2}
\eeq
 The metric $g^{\mu\nu}$ depends on only $x$. Constraints
 are $p_\lambda\equiv\pi\approx 0$ and $p_y\approx 0$ as before.\par
   Since the Hamiltonian of the reduced phase space quantization; 
\beq
             H=\frac{1}{2}\gamma^{ij}p_i p_j   \label{eqn:h3}
\eeq
has a non-trivial metric in this case, the Hamiltonian operator is
\beq
  \hat{H}=\frac{1}{2\sqrt{\gamma}}\partial_i\sqrt{\gamma}\gamma^{ij}
          \partial_j + F(R,\cdots).
\eeq
Here $F$ is a function of $R,\cdots $of $\gamma_{ij}$. In this model the
reduced phase space quantization may have an additional function F, too.\par
 While the Hamiltonian operator of Dirac quantization is
\beqa
  \hat{H}&=&\frac{1}{2\sqrt{g}}\partial_\mu\sqrt{g}g^{\mu\nu}
          \partial_\nu + G(R,\cdots)\nonumber\\
        &=&\frac{1}{2M\sqrt{\gamma}}\partial_i\sqrt{\gamma}M\gamma^{ij}
  \partial_j-\frac{1}{2M\sqrt{\gamma}}\partial_i\sqrt{\gamma}N^i\partial_y
 -\frac{1}{2M\sqrt{\gamma}}\partial_y\sqrt{\gamma}N^j\partial_j\nonumber\\
&+&\frac{1}{2M\sqrt{\gamma}}\partial_y\sqrt{\gamma}\frac{N^2}{M}\partial_y
   +\frac{1}{2M\sqrt{\gamma}}\partial_y\sqrt{\gamma}\partial_\pi
+\frac{1}{2M\sqrt{\gamma}}\partial_\pi\sqrt{\gamma}\partial_y \nonumber\\
  &+&G(R,\cdots).
\eeqa
Here $G$ is a function of $R,\cdots$ of $g_{\mu\nu}$. The constraint
 $\hat{p}_y=\partial_y\approx 0$ makes the Hamiltonian
\beq
     \hat{H}=\frac{1}{2M\sqrt{\gamma}}\partial_i\sqrt{\gamma}M\gamma^{ij}
   \partial_j + G. 
\eeq
\par
    The inner product for the reduced phase quantization is defined as
\beq
             \int \sqrt{\gamma}dx\Psi_r^* \Psi_r.
\eeq
On the other hand, for Dirac quantization, it is written by
\bdm
            \int \sqrt{\gamma}M dxdyd\pi \Psi_D^*\Psi_D
\edm
\beq
           =\int dy \int \sqrt{\gamma}M dx \Psi_D^*\Psi_D
\eeq
as before. For the both inner products to agree with each other
\beq
                    \Psi_D=\frac{1}{\sqrt{\gamma}}\Psi_r
\eeq
must be satisfied in this case.\par
 The expectation value of the energy for the reduced phase space
 quantization is
\beq
 <E>_r=\int \sqrt{\gamma} dx \Psi_r^*(\frac{1}{2\gamma}\partial_i
\sqrt{\gamma}\gamma^{ij} \partial_j+F)\Psi_r.
\eeq
While in Dirac quantization it is given by
\beqa
 <E>_D&=&\int \sqrt{\gamma}Mdx \frac{1}{\sqrt{M}}\Psi_r^*
(\frac{1}{2\gamma}\partial_i \sqrt{\gamma}\gamma^{ij} \partial_j+G)
  \frac{1}{\sqrt{M}} \Psi_r\nonumber\\
      &=&<E>_r\nonumber\\
       &+&\int\sqrt{\gamma} dx \Psi^*(-\frac{1}{2}\frac{\gamma^{ij}
\nabla_i\nabla_j\sqrt{M}}{\sqrt{M}}+G-F)\Psi_r \label{eqn:e1}
\eeqa
where $\nabla_i$ is a covariant derivative with respect to $\gamma_{ij}$.
For the both quantization to be equivalent, the second term should be 
zero in the second equation.\par
  To simplify the problem, suppose that $\gamma_{ij}$ is a two-dimensional
metric. The dimension of the space on which Dirac quantization is
 performed is four. Four-dimensional $R, R^{\mu\nu}R_{\mu\nu}$, and 
$R^{\mu\nu\lambda\sigma}R_{\mu\nu\lambda\sigma}$ of the metric(\ref{eqn:g1})
 and (\ref{eqn:g2}) are related with two- dimensional ones of the metric
 $\gamma_{ij}$ as
\beq
  R=R^{(2)}-4\gamma^{z\bar{z}}\frac{\nabla_z \nabla_{\bar{z}}\sqrt{M}}
      {\sqrt{M}} -2\frac{\gamma^{z\bar{z}}\nabla_z \nabla_{\bar{z}}M}{M},
      \label{eqn:r1}
\eeq
\beqa
    R^{\mu\nu}R_{\mu\nu}&=& R^{(2)ij}R_{ij}^{(2)} 
     -4R^{z\bar{z}}\frac{2\nabla_z\nabla_{\bar{z}}\sqrt{M}}{\sqrt{M}}
+2(2\gamma^{z\bar{z}}\frac{\nabla_z\nabla_{\bar{z}}\sqrt{M}}
{\sqrt{M}})^2\nonumber\\
&+&8(\gamma^{z\bar{z}})^2\frac{\nabla_z \nabla_z \sqrt{M}}{\sqrt{M}}\frac{
  \nabla_{\bar{z}} \nabla_{\bar{z}}\sqrt{M}}{\sqrt{M}}
  +\frac{1}{2}(\gamma^{z\bar{z}}\frac{2\nabla_z \nabla_{\bar{z}}M}{M})^2,
      \label{eqn:r2}
\eeqa
\beqa
R^{\mu\nu\lambda\sigma}R_{\mu\nu\lambda\sigma}&=& R^{(2)ijkl}R_{ijkl}^{(2)}+
4(2\gamma^{z\bar{z}}\frac{\nabla_z\nabla_{\bar{z}}}{\sqrt{M}}{\sqrt{M}})^2
\nonumber\\
 &+&16(\gamma^{z\bar{z}})^2\frac{\nabla_z \nabla_z \sqrt{M}}{\sqrt{M}}
\frac{\nabla_{\bar{z}} \nabla_{\bar{z}}\sqrt{M}}{\sqrt{M}}\nonumber\\
& +&(4\gamma^{z\bar{z}}\frac{\nabla_z\nabla_{\bar{z}}\sqrt{M}}{\sqrt{M}}-
  \gamma^{z\bar{z}}\frac{2\nabla_z\nabla_{\bar{z}}M}{M})^2,
      \label{eqn:r3}
\eeqa
where $R^{(2)}, R^{(2)}_{ij}$, and$ R^{(2)}_{ijkl}$ are two-dimensional ones.
 We use a complex coordinate
in two-dimension where $\gamma_{z\bar{z}}\not= 0$ and $\gamma_{zz}=\gamma_
{\bar{z}\bar{z}}=0$. Using relations $R^{ij}R_{ij}=\frac{1}{2}R^2$,
 $R^{ijkl}R_{ijkl}=R^2$, and $R^{z\bar{z}}=\frac{1}{2}R\gamma^{z\bar{z}}$
 in two-dimension and defining
\beq
          a\equiv
  2\gamma^{z\bar{z}}\frac{\nabla_z\nabla_{\bar{z}}\sqrt{M}}{\sqrt{M}},
\eeq
\beq
         b\equiv
  (\gamma^{z\bar{z}})^2\frac{\nabla_z \nabla_z \sqrt{M}}{\sqrt{M}}\frac{
  \nabla_{\bar{z}} \nabla_{\bar{z}}\sqrt{M}}{\sqrt{M}},
\eeq
\beq
          B\equiv
  2\gamma^{z\bar{z}}\frac{\nabla_z\nabla_{\bar{z}}M}{M},
\eeq
we can rewrite equations (\ref{eqn:r1}),(\ref{eqn:r2}), and (\ref{eqn:r3}) as
\beq
      R=R^{(2)}-2a-B,
\eeq
\beq
 R^{\mu\nu}R_{\mu\nu}=\frac{1}{2}R^{(2)2}-2R^{(2)}a+2a^2+8b+\frac{B^2}{2},
\eeq
\beq
   R^{\mu\nu\lambda\sigma}R_{\mu\nu\lambda\sigma}= R^{(2)2}+4a^2+16b
   +(2a-B)^2.
\eeq
From these equations, we get
\beq
         \frac{a}{2}=\frac{-R\pm\sqrt{R^2-6R^{\mu\nu}R_{\mu\nu}+
              3R^{\mu\nu\lambda\sigma}R_{\mu\nu\lambda\sigma}}}{12}.
             \label{eqn:f1}
\eeq
Therefore, if we take this quantity as G in equation (\ref{eqn:e1}), the
 reduced phase space quantization coincide with Dirac quantization. Since
 two-dimensional quantity does not appear in the right hand side of
 equation (\ref {eqn:f1}), F in equation(\ref{eqn:e1}) is zero in two-
dimension.\par
Now we get two operator orderings. The Hamiltonian operator for the
 equation (\ref {eqn:h2}) in four-dimension is
\beq
     \hat{H}=\frac{1}{2}\bigtriangleup+
             \frac{-R+\sqrt{R^2-6R^{\mu\nu}R_{\mu\nu}+
              3R^{\mu\nu\lambda\sigma}R_{\mu\nu\lambda\sigma}}}{12},
             \label{eqn:o2}
\eeq
where we take positive sign of root as before. And for the two-dimensional
Hamiltonian of equation (\ref{eqn:h3}), the Hamiltonian operator is
\beq
     \hat{H}=\frac{1}{2}\bigtriangleup.
      \label{eqn:o3}
\eeq
In two-dimension there does not appear any function of $R$.\par
 So far we get three operator orderings. These operator orderings have the
relations each other. In the three-dimensional constraint system of the
metric of equations (\ref{eqn:g1}) and (\ref{eqn:g2}),
$R^{\mu\nu\lambda\sigma}R_{\mu\nu\lambda\sigma}$ is written as
\beq
R^{\mu\nu\lambda\sigma}R_{\mu\nu\lambda\sigma}=4R^{\mu\nu}R_{\mu\nu}-R^2
\eeq
Substituting this equation into equation (\ref{eqn:o2}), we get equation
(\ref{eqn:o1}). 
In two-dimension $ R^{\mu\nu}R_{\mu\nu}=\frac{1}{2}R^2$ is satisfied. If
we substitute this relation into equation ({\ref{eqn:o1}), we get two-
dimensional trivial Hamiltonian operator (\ref{eqn:o3}). This is the
 reason why we take the positive sign of root.

\section{Conclusion and discussion}
We showed the equivalence of the reduced phase space quantization and
 Dirac quantization. Both methods are the different operator formalism
out of the same path integral. Using this equivalence and the 
reparametrization invariance, we determined operator orderings in three 
examples. However, these expressions are not unique.  Because scalars are
 expressible by other scalars. We can derive many equivalent forms.\par
  In general the n-dimensional Hamiltonian operator is determined by the
 equivalence with the artificially extended (n+2)-dimensional constraint
 system and the (n+2)-dimensional Hamiltonian operator of the constraint 
system is determined at the same time. However, it is difficult to
determine the concrete form of the quantum potential.\par
 In the case of the quantum gravity, the Hamiltonian operator is not
positive definite. However, this method is applicable to the quantum
 gravity. For example, in the minisuperspace model with scale
factor and scalar matter,  Weeler-DeWitt equation reduces $\Box\Psi=0$.
Because in the case of two-dimension there does not appeare the quantum
potential.
  
\acknowledgments
I am grateful to A. Fujitsu for valuable discussion.


\begin{thebibliography}{}
\bibitem{kun} A.Ashteker and G.T.Horowitz {it\ Phys. Rev.}{\bf D26},
              3342 (1982);
              K.V.Kuchar {\it Phys. Rev.}{\bf D34}, 3031 (1986);
              {\it Phys. Rev.}{bf D34}, 3044 (1986);
              J.D.Romano and R.S.Tate {\it Class. Quantum Grav.}
              {\bf 6}, 1487 (1989);
              K.Schleich {\it Class. Quantum Grav.}{\bf 7}, 1529 (1990);
              G.Kunstatter {\it Class. Quantum Grav.}{\bf 9}, 1469 (1992); 
              {\it GENERAL RELATIVITY AND RELATIVISTIC
               ASTROPHYSICS} (Singapore: World Scientific) p~171 (1992)\\
\bibitem{shi} K.Shimizu and S.Wada {\it Int. J. Mod. Phys. A}
              {\bf Vol.7 No.8}, 1627 (1992)\\
\bibitem{fra}E.S.Fradkin and G.A.Vilkovisky {\it CERN Report No. TH2332}
             (1977); M.Henneaux {\it Phys. rep.}{\bf 126}, 1 (1985)
\end{thebibliography}
 \end{document}